\documentclass{appolb}
\usepackage{epsfig}

\begin{document}

\title{Study of Deformation Effects \\ in the Charged Particle Emission from
$^{46}$Ti$^{*}$ \thanks{Presented at the XXXIX Zakopane School of Physics -
International Symposium "Atomic nuclei at extreme values of temperature,
spin and isospin" - Zakopane, Poland, August 31 - September 5, 2004}}

\author{M.~Brekiesz$^a$, P.~Papka$^{b,c}$, A.~Maj$^a$, M.~Kmiecik$^a$, C.~Beck$^b$,
P.~Bednarczyk$^{a, d}$, J.~Gr\c{e}bosz$^a$, F. ~Haas$^b$,
W.~M\c{e}czy\'nski$^a$, V.~Rauch$^b$, M.~Rousseau$^b$, A.~S\`{a}nchez~i~Zafra$^b$,
J.~Stycze\'n$^a$, 
S.~Thummerer$^e$, M.~Zi\c{e}bli\'nski$^a$ and K.~Zuber$^a$
\address{
$^a$The Niewodnicza\'nski Institute of Nuclear Physics, PAN, Krak\'ow, Poland \\
$^b$IReS IN$_{2}$P$_{3}$-CNRS/Universit\'e Louis Pasteur,
         Strasbourg, France\\
$^c$Department of Physics, University of York, Heslington, United
Kingdom\\
$^d$Gesellschaft f\"ur Schwerionenforschung, Darmstadt, Germany\\
$^e$Hahn-Meitner Institute, Berlin, Germany}
}

\maketitle

\begin{abstract}

The $^{46}$Ti$^{*}$ compound nucleus, as populated by the fusion-evaporation
reaction $^{27}$Al + $^{19}$F at the bombarding energy of 144~MeV, has been
investigated by charged particle spectroscopy using the multidetector array
{\sc ICARE} at the {\sc VIVITRON} tandem facility of the IReS (Strasbourg).
The light charged particles have been measured in coincidence
with evaporation residues. The {\sc CACARIZO} code, a Monte Carlo
implementation of the statistical-model code {\sc CASCADE}, has been used
to calculate the spectral shapes of evaporated $\alpha$-particles which are
compared with the experimental spectra. This comparison indicates the possible
signature of large deformations of the compound nucleus.

\end{abstract}

\PACS{25.70.Gh, 25.70.Jj, 25.70.Mn, 24.60.Dr}

\section{Introduction}

In the recent years, there have been a number of experimental and theoretical
studies aimed at understanding the effects of large deformations in the case of
light-mass nuclei. The very elongated prolate or triaxial shapes were deduced
from the spectra of the Giant Dipole Resonance (GDR) from the decay of
$^{46}$Ti$^{*}$~\cite{Maj01,Maj02} and $^{45}$Sc$^{*}$ ~\cite{Kic03}. The
results were consistent with predictions of the LSD (Lublin-Strasbourg Drop)
model~\cite{Pom06,Dub07}, in which the large deformations are ascribed to the
Jacobi shape transition. The large deformations were also indicated by the
measurement of energy spectra and angular distributions of the 
light charged particles (LCP) originated from the decay 
of $^{44}$Ti$^{*}$ \cite{Pap04b} as formed in two fusion reactions
$^{16}$O + $^{28}$Si~\cite{Pap03} and $^{32}$S + $^{12}$C~\cite{Pap04a}. In
addition, a number of superdeformed bands of discrete $\gamma$-ray transitions were
discovered in selected $N~=~Z$ nuclei belonging to this mass region
(e.g.~\cite{Ide,Beck04}).
In this paper, we focus on the measurement of the LCP spectra in coincidence
with evaporation residues (ER) for the reaction $^{27}$Al~+~$^{19}$F at a bombarding
energy of $E_{lab}$($^{27}$Al)~=~144~MeV. In the following sections the
experimental setup, the data analysis and the discussion of the
preliminary results are presented.

\section{The experimental setup and experimental results}

The experiment was performed at the {\sc VIVITRON} tandem facility
of the IReS Strasbourg (France), using the multidetector array {\sc
ICARE} and a large volume (4$^"\times$4$^"$) BGO detector. 
The $^{46}$Ti$^{*}$ compound nucleus (CN)
was populated by the $^{27}$Al~+~$^{19}$F reaction at the 144~MeV bombarding
energy of the aluminium beam. The inverse kinematics reaction was chosen to
increase the residual nuclei velocity in order to resolve the highest $Z$ 
(the slowest evaporation residues) produced in the reaction.
 
A fluorine target (LiF: 150 $\mu$g/cm$^2$ of F, 55 $\mu$g/cm$^2$ of Li) 
on a thin carbon backing (20 $\mu$g/cm$^2$) was used. 
The excitation energy of the $^{46}$Ti nuclei was
85~MeV and the maximum angular momentum $L_{crit}\approx~$35~$\hbar$.

High-energy $\gamma$-rays from the GDR decay were measured
using the BGO detector. The heavy fragments were detected in six
gas-silicon telescopes (IC), each composed of a 4.8~cm long ionization
chamber with a thin Mylar entrance window followed by a 500~$\mu$m thick
surface barrier silicon detector. 
The IC were located at $\Theta_{lab}=\pm 10^{\circ}$ in three
distinct reaction planes. The in-plane detection of coincident LCP's was done
using ten triple telescopes (40~$\mu$m Si, 300~$\mu$m Si, 2~cm CsI(Tl)) placed
at forward angles 
($\Theta_{lab}=\pm 20^{\circ},\pm 25^{\circ},\pm 30^{\circ}, \pm 35^{\circ},
\pm 40^{\circ}$), eighteen two-element telescopes (40~$\mu$m Si, 2~cm CsI(Tl))
placed at forward and backward angles ($\pm 45^{\circ}~\leq~\Theta_{lab}~\leq\pm
130^{\circ}$) with $\Delta\Theta=5^{\circ}$ angular step, and finally four other IC
telescopes located at the most backward angles 
($ \pm150^{\circ}~\leq~\Theta_{lab}~\leq\pm 165^{\circ}$).
The opening angle of the detectors was $\approx3^{\circ}$.
The IC were filled with isobutane
at a pressure of 40.5~mbar for the forward angles and of 49.6~mbar in
backward angles, thus allowing for simultaneous measurement of both heavy and
light fragments.

The energy calibrations of various telescopes of the {\sc
ICARE} multidetector array were done using radioactive $^{228}$Th  and
$^{241}$Am $\alpha$-particle sources in the \mbox{5-9}~MeV energy range and elastic
scattering of 45~MeV $^{11}$B, 53~MeV $^{16}$O and 144~MeV $^{27}$Al from
$^{197}$Au target. In addition, the 
$^{12}$C($^{16}$O,$\alpha$)$^{24}$Mg$^{*}$ reaction  at 53~MeV was used to provide
known energies of $\alpha$-particles feeding the $^{24}$Mg excited states, thus
allowing for an accurate calibration of the backward angle detectors.

Exclusive energy spectra of the $\alpha$-particles emitted 
in the laboratory frame at the angles $\Theta_{lab}~=~25^{\circ}, 35^{\circ}, 45^{\circ}$
 in coincidence with $Z$~=~18, 19, 20
measured by ionization chamber located at $\Theta_{lab}=~10^{\circ}$ are
shown in Fig.~1 by the solid points. All measured spectra
have the expected Maxwellian shape with an exponential fall-off at high energy.
The dashed and solid lines are the results of statistical-model calculations.

\begin{figure}[htb]
\begin{center}
\epsfxsize=125mm
\epsfbox{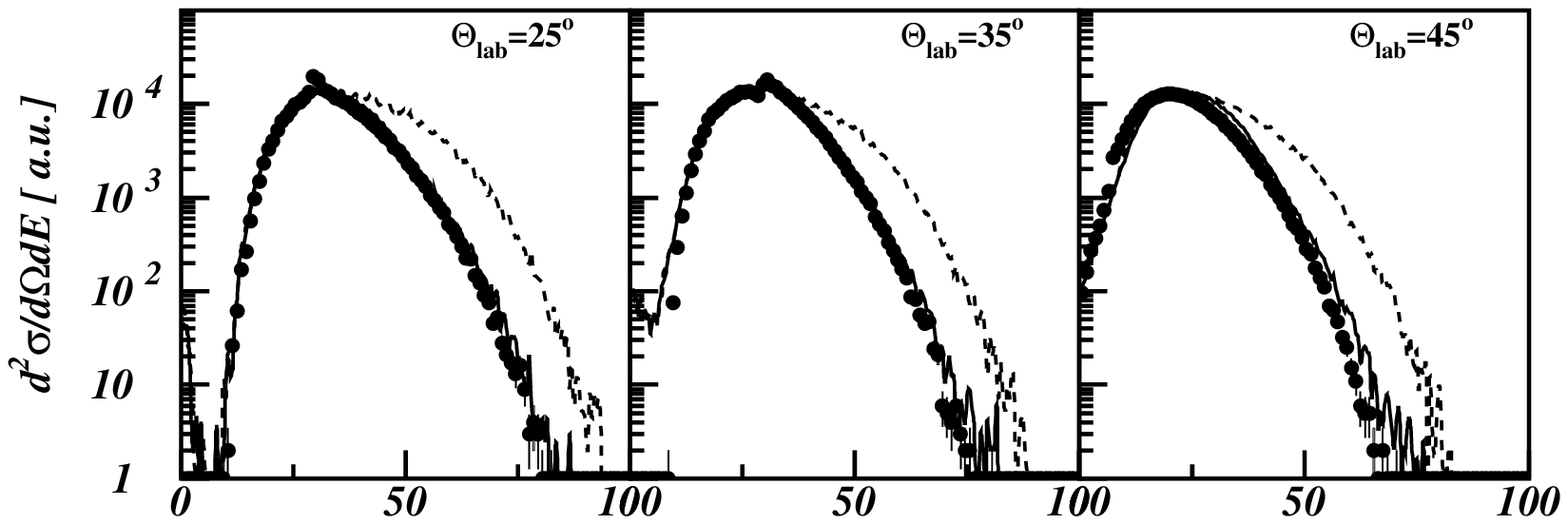}\\
\epsfxsize=125mm
\epsfbox{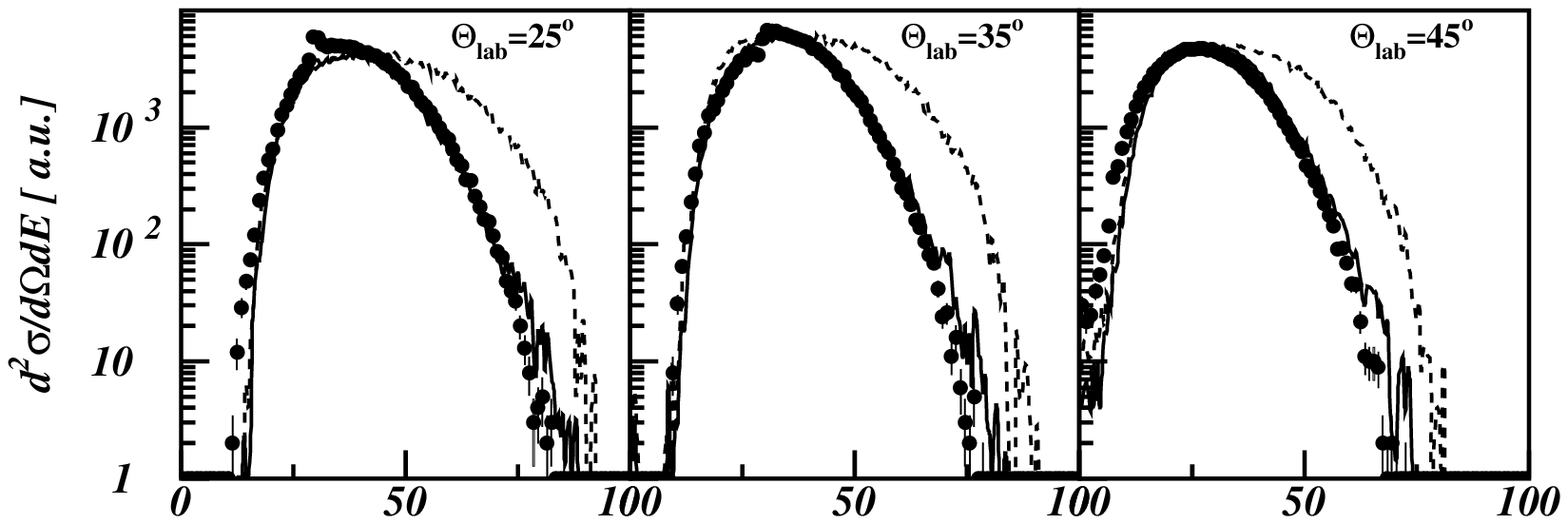}\\
\epsfxsize=125mm
\epsfbox{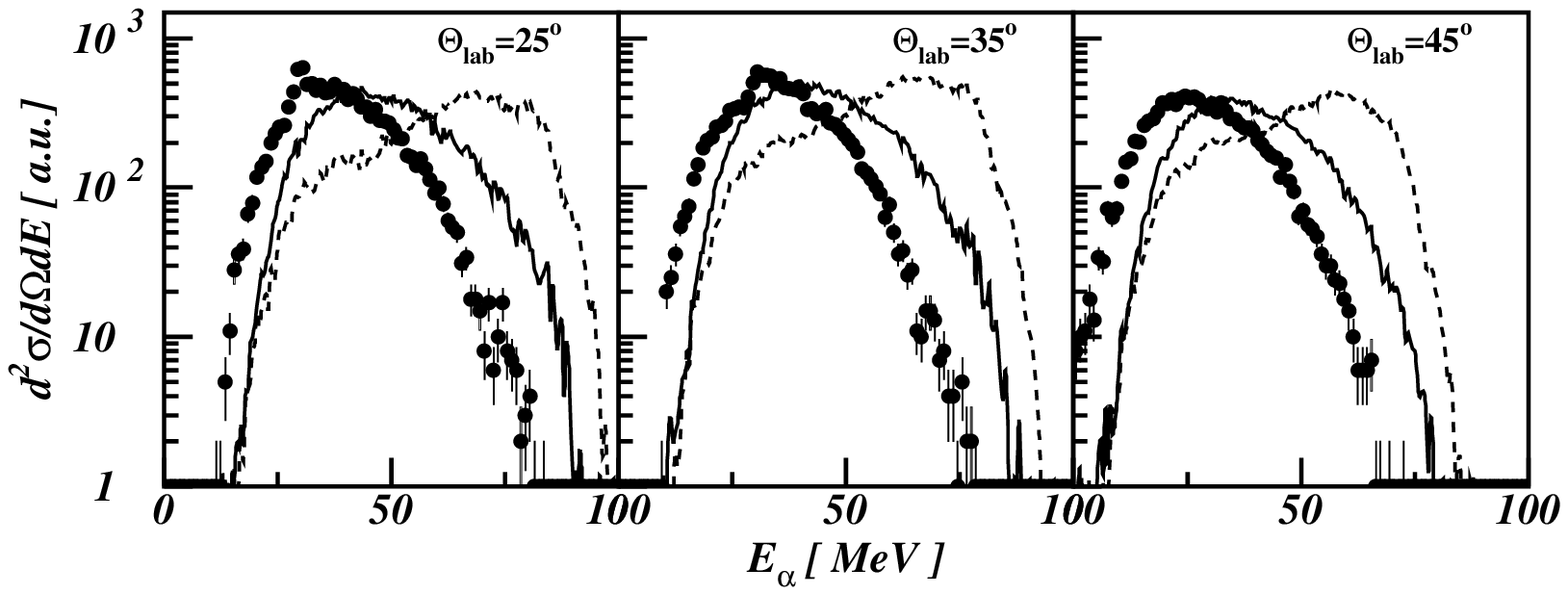}
\caption{Experimental (points) and calculated (lines) $\alpha$-particle energy spectra for
$\Theta_{lab}=25^{\circ},35^{\circ},45^{\circ}$
in coincidence with three different ER's ($Z$~=~18: upper row, $Z$~=~19: middle row,
$Z$~=~20: bottom row) detected in IC at $\Theta_{lab}~=~10^{\circ}$. 
The calculations were carried out with deformation
parameters for RLDM (dashed line) and for very elongated prolate (solid
line) shapes (see text).}
\end{center}
\end{figure}

\section{Statistical-model calculations and discussion }

The analysis of LCP data has been performed using {\sc CACARIZO}, the Monte
Carlo version of the statistical-model code {\sc CASCADE} ~\cite{Pul08}, which
is based on the Hauser-Feshbach formalism ~\cite{Bha09,Rou10,Mahboub}.
The compound nucleus decays are followed step-by-step and evaporated LCP's 
and neutrons are recorded in an event file.
The main practical advantage of the code is that the effective experimental 
geometry (solid angle and position)
of the {\sc ICARE} detectors is taken into account in the calculation of
the charged particle energy spectra. Therefore one can, on the event-by-event basis,
convert the calculated energies into the laboratory frame for an easier 
comparison of the calculated and measured spectra.

The sensitivity of various parameters for the statistical description,
as the nuclear level densities and barrier transmission probabilities,
has been discussed in detail in Ref.~\cite{Mahboub}. In the
calculations, the transmission coefficients $T_{l}$ of all competing
evaporation channels have been evaluated with optical model parameters for
spherical nuclei~\cite{Hui11}. The choice of the transmission coefficients 
is of particular importance in the near-barrier region, 
where they define the emission probability of low-energy particles.
Above the barrier, the kinetic energy of LCP is higher than 
the potential barrier and the choice of $T_{l}$ parameterizations is less
important (see Ref. ~\cite{Mahboub}). 
The high-energy part of $\alpha$-particle spectra depends 
on the available phase space, which is
obtained  by the statistical model from the spin-dependent level density. The
level density is calculated using the Rotating Liquid Drop Model (RLDM)
~\cite{Pul08} and can be changed using the deformability parameters.
Larger deformations lower the yrast line, what increases the level density at
higher available excitation energy of the final nucleus after $\alpha$
emission, thus reduce $\alpha$-particle energies.
In the code, the yrast line is parameterized with deformability parameters 
$\delta_{1}$ and $\delta_{2}$: $E_{L}=\hbar^{2}L(L+1)/2\Im_{eff}$ 
with $\Im_{eff}=\Im_{sphere}(1+\delta_{1}L^{2}+\delta_{2}L^{4})$
~\cite{Rou10}, where $\Im_{eff}$ is the effective moment of 
inertia, $\Im_{sphere}$~=~(2/5)~$A^{5/3}r_{0}^{2}$
 is the rigid body moment of inertia 
of the spherical nucleus and $r_{0}$ is the radius parameter 
(set to 1.3 fm in the present calculations).

For the calculations, two different deformation sets are applied 
with different yrast line shape parameterizations.
Fig.~2 illustrates the yrast line for a spherical rigid body 
and the two yrast lines used in present calculations: 
for shapes following the RLDM predictions 
(spherical up to $L$~=~24~$\hbar$ and nearly spherical above) 
and for very elongated prolate shapes 
($\delta_{1}$~=~4.7$\times10^{-4}$, $\delta_{2}$~=~1$\times10^{-7}$).
The latter values of the deformability parameters are taken from
the previous studies on the $^{44}$Ti nucleus \cite{Pap04b,Pap03,Pap04a}.
The angular momentum distribution used in the statistical-model calculation is
dependent on the diffusivity parameter $\Delta$$L$ and the critical angular
momentum for fusion $L_{crit}$. For the final calculations the value of 
$\Delta$$L$~=~1 is used
~\cite{Bha09,Rou10,Mahboub} and the value of $L_{crit}~=~$35~$\hbar$ is 
deduced from the fusion cross section data~\cite{Zin12}. 

\begin{figure}[htb]
\begin{center}
\epsfig{file=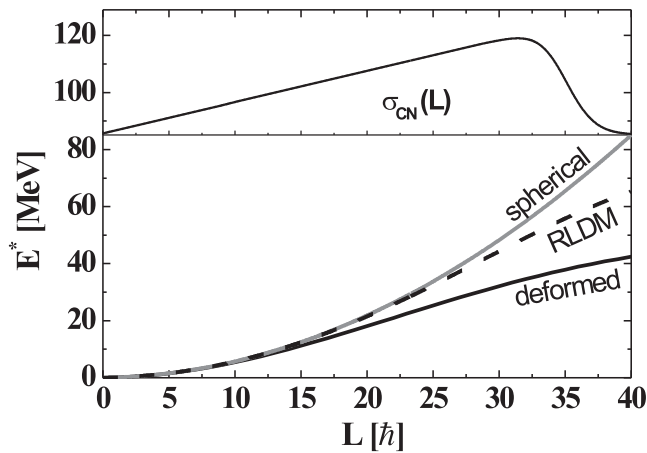,height=6.0cm}
\caption{The yrast lines for spherical rigid body, for shapes
predicted by the RLDM and for very elongated prolate shapes (see text)
of the $^{46}$Ti compound nucleus. The assumed angular 
momentum distribution of the compound nucleus is shown on the top.}

\end{center}
\end{figure}

The dashed lines in Fig.~1 show the energy spectra of $\alpha$-particles 
predicted by {\sc CACARIZO} assuming the RLDM yrast line. It can be observed 
that average energies of $\alpha$-particles measured in coincidence with all ER
($Z$~=~18, 19, 20) are systematically lower than predicted.

The solid lines in Fig.~1 illustrate
the calculations of {\sc CACARIZO}, which were obtained using the yrast line
deformability parameters corresponding to large prolate deformations (such yrast
line corresponds, on average, to the yrast line of a rigid body with the
deformation parameter $\beta\approx 0.6$). 
As can be seen, such parameterization results in a very good 
overall reproduction of the spectral shapes of the
$\alpha$-particles detected in coincidence either with $Z$~=~18 or with $Z$~=~19.

However, the spectra associated with $Z$~=~20 are in disagreement with
the calculations. In order to improve the agreement, an even more deformed yrast
line (rather unrealistic) would be required. One should note here, that the
condition for the $\alpha$-particle and the $Z$~=~20 residue to belong to the
same event induces a severe narrowing of the available phase space to only 
the highest angular momenta of CN. 
This might suggest indeed the occurence of very elongated
shapes around $L$~=~30~$\hbar$, as expected for the Jacobi shape transition, 
confirming the GDR results of Ref.~\cite{Maj02}.

However, a more refined analysis, including also a consistent treatment of 
the energy spectra of the protons, the LCP angular correlations 
and the high-energy $\gamma$-ray spectra, 
will have to be undertaken in order to confirm such hypothesis.

This work was partially supported by the Polish Committee for Scientific
Research (KBN Grant No. 2 P03B 118 22) and by the exchange programme between
the {\em Institut National de Physique Nucl\'eaire et de Physique des
Particules, $IN_2P_3$}, and Polish Nuclear Physics Laboratories.
The authors wish also to thank the staff of the {\sc VIVITRON} for providing
us with a good quality $^{27}$Al beam, M.A. Saettel for providing the targets,
and J. Devin and C. Fuchs for the excellent support in carrying out this
experiment.

\end{document}